\documentclass{moriond}

\def\Journal#1#2#3#4{{#1} {\bf #2}, #3 (#4)}

\def\PLB{{\em Phys. Lett.}  B}
\def\PRL{\em Phys. Rev. Lett.}
\def\PRD{{\em Phys. Rev.} D}

\def\ttbar{\ensuremath{\mathrm t\overline{\mathrm{t}}}}
\def\GeV{\ensuremath{\:\mathrm{GeV}}}

\newcommand{\Photo}{\includegraphics[height=35mm]{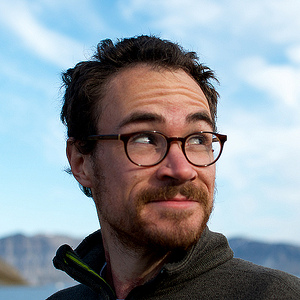}}

\begin{document}

\vspace*{4cm}
\title{NEW METHODS FOR TOP-QUARK MASS MEASUREMENTS AT THE LHC}

\author{B. STIEGER (University of Nebraska, Lincoln), \\ \em on behalf of the ATLAS and CMS collaborations}

\address{~}

\maketitle\abstracts{
Several recent new measurements of the top-quark mass that use alternative observables and reconstruction techniques are presented, performed by the ATLAS and CMS collaborations at the CERN LHC.\@
Alternative observables can help provide insight by presenting different systematic sensitivities and by constraining prevailing systematic uncertainties of standard measurements, such as jet energy calibrations.
Furthermore, the top-quark mass is extracted from theoretically well-defined observables, such as the inclusive production cross section for top quark pairs.
Finally, the mass is measured in event topologies dominated by  electroweak-mediated single top production by both experiments.
The results of different techniques and production modes are found to be consistent with what is obtained in standard measurements.
}

\section{Introduction}
The mass of the top quark is a fundamental parameter of the standard model (SM) and---together with the masses of the W and Higgs bosons and other SM parameters---provides a strong self-consistency check of the theory.
Furthermore, the value of the top-quark mass has a significant impact on the evolution of the Higgs quartic coupling, affecting the overall stability of the electroweak vacuum.
A top quark heavier by a few \GeV\ leads to a prediction of an unstable vacuum already many orders of magnitude below the Planck scale, indicating the presence of physics beyond the standard model at such a scale.

The top quark has been studied in great detail since experimentally establishing its existence more than 20 years ago at the Tevatron collider.
Its mass has since been measured with ever-increasing accuracy, using methods that attempt a full kinematic reconstruction of the \ttbar\ final state.
The most precise measurements exploit several observables in a multidimensional fit to constrain the leading sources of experimental uncertainties.
Currently the world's best measurement~\cite{cmslegacy}, by the CMS collaboration, yields a value of $172.44 \pm 0.13 (\mathrm{stat.}) \pm 0.47 (\mathrm{syst.}) \GeV$, i.e.\ a precision of just below three per mill and in good agreement with the 2014 world average~\cite{worldcomb,cmsdet,atlasdet}.
The overall precision of such measurements is limited by our understanding of the modeling of b-quark hadronization in the used simulations.

Complementary measurements can help further improve the overall precision in combination with standard methods by using observables that are less dependent on the modeling of hadronization, e.g.\ by avoiding the use of jet kinematics.
Finally, the top-quark mass can be obtained by precisely measuring production cross sections---either inclusively or differentially in bins of a kinematic observable sensitive to the mass.
This allows the extraction of the mass as a well-defined parameter in the context of a renormalization scheme of quantum chromodynamics (QCD).

\section{Measurements without jets}\label{sec:expclean}
The heavy impact of hadronization modeling uncertainties in most top-quark mass measurements stems from the usage of jet kinematics in the mass-sensitive observables.
Using kinematic properties of \ttbar\ events without using jets often relegates the issue to merely affecting the acceptance of collision events and playing a minor role in the determination of the top-quark mass.

\subsection{Using secondary vertices and leptons}\label{subsec:secvtx}
A recently published analysis by the CMS experiment\cite{cmssectvx} exploits an observable using secondary vertices reconstructed in b-quark initiated jets, i.e.\ relying on the high momentum resolution for charged particles reconstructed in the tracking detector.
The observable combines these secondary vertices with a charged lepton from the W-boson decay to form an invariant mass that is highly sensitive to the top-quark mass.
The analysis is performed in exclusive bins of the multiplicity of tracks used in the secondary vertex reconstruction (exactly 3, 4, or 5 tracks) and makes use of events both with one or two charged leptons.
All (up to four) possible lepton-vertex combinations in each event are used.

The analysis yields a top-quark mass of $173.68 \pm 0.2 (\mathrm{stat}) {}^{+1.58}_{-0.97} (\mathrm{syst})\GeV$, calibrated on simulated events.
The leading source of systematic uncertainty is the modeling of b-quark fragmentation in the simulation, i.e.\ the fraction of b-parton momentum transfered to the b hadron.
Furthermore, the analysis is sensitive to the modeling of the top quark transverse momentum which has previously been observed to be inadequately described by simulation.
Experimental sources of systematic uncertainty and the overall modeling of hadronization do not have a strong impact on the analysis.

\begin{figure}
\hfill
\begin{minipage}{0.48\linewidth}
\centerline{\includegraphics[width=1.0\linewidth]{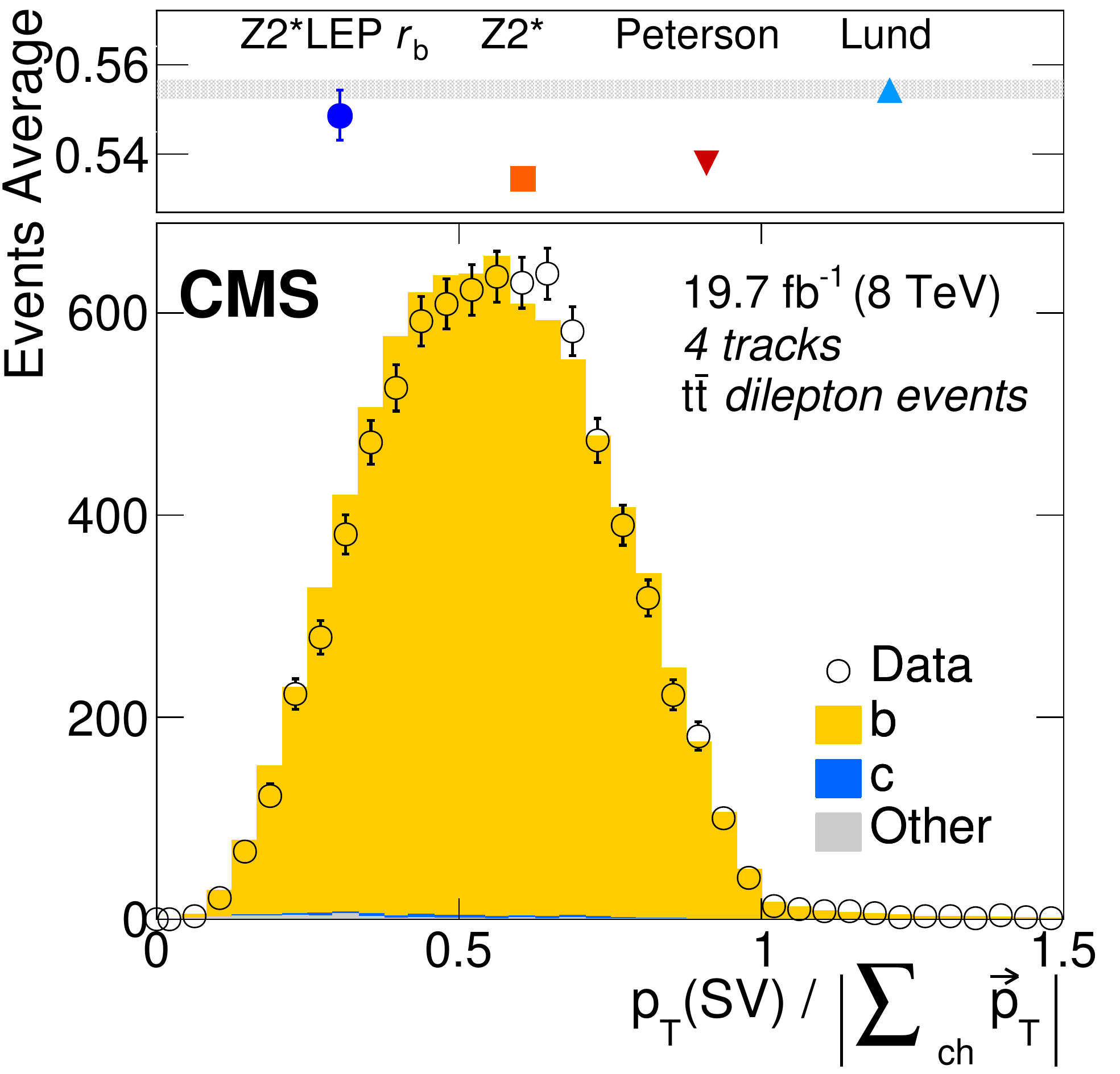}}
\end{minipage}
\hfill
\begin{minipage}{0.49\linewidth}
\centerline{\includegraphics[width=1.0\linewidth]{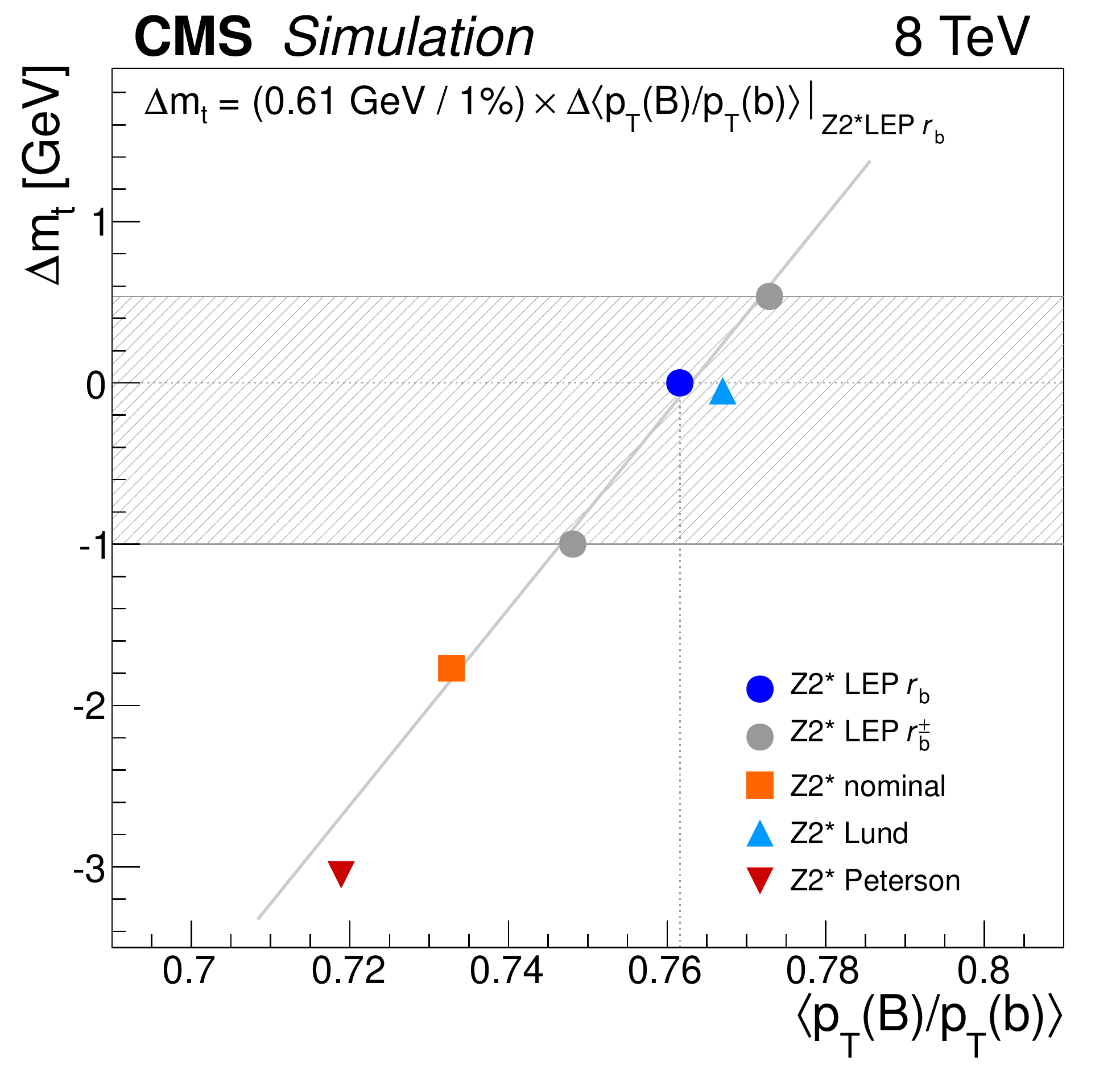}}
\end{minipage}
\caption[]{
  Left: Fraction of the overall transverse momentum of charged particles clustered in a jet carried by the reconstructed secondary vertex in dileptonic \ttbar\ events~\cite{cmssectvx}. The simulation (filled histogram) using the Z2*LEP$r_{\mathrm{b}}$ tune is compared to the observed data distribution (open markers). The top panel compares the average of the distribution in the data (grey band) to various b-quark fragmentation models.
  Right: Impact of a change in average b parton-to-hadron momentum transfer on the extracted top-quark mass~\cite{cmssectvx}. The extracted mass is shown for the average momentum transfer values of the same b-quark fragmentation models. A linear fit (grey line) shows a change of $0.61\GeV$ for each percent change of momentum fraction.
}
\label{fig1}
\end{figure}

The paper includes a study comparing different b fragmentation models with the observed data, using the ratio of the overall transverse momentum of charged particles clustered in a jet and the transverse momentum carried by the secondary vertex, see Fig.~\ref{fig1} (left).
The impact of a change in the average parton-to-hadron momentum transfer on the extracted central value of the top-quark mass is shown in Fig.~\ref{fig1} (right), where a change of average momentum of one percent leads to a shift in the top-quark mass of about $0.6\GeV$.

\subsection{Using charm mesons and leptons}\label{subsec:charm}
An experimentally even cleaner observable can be obtained by exclusively reconstructing charm mesons in the b-hadron decay and pairing them with a lepton from the W-boson decays.
The idea of such a measurement, using J$/\Psi$ mesons, has been proposed already during LEP times~\cite{jpsitheo}, and was now implemented for the first time by the CMS collaboration using 8~TeV data from the 2012 LHC run~\cite{cmsjpsi}.
With the current amount of data, the study is statistically limited, but an analysis of systematic uncertainties shows promising prospects.
The relevant experimental precision on the top-quark mass is found to be better than 100~MeV.
The limiting factor of the current analysis lies in the uncertainty in the modeling of the top quark transverse momentum and in the effect of QCD scale variations.
It is worth to note that the lepton-J$/\Psi$ invariant mass shows less sensitivity---roughly half the effect---to the modeling of the b quark fragmentation than the lepton-secondary vertex mass of the previous analysis.

\subsection{Using lepton kinematics}\label{subsec:lepkin}
Another possibility to avoid hadronization-related uncertainties is to only use leptonic observables.
A recent proposal~\cite{frixmitov} to measure the top-quark mass from the kinematical properties of the dilepton system in \ttbar\ events was recently implemented by CMS~\cite{lepkin}.
While experimentally only limited by the lepton momentum scale calibration, the analysis is hampered by uncertainties from scale variations in the signal MC and the modeling of top quark kinematics.
However, the use of next-to-leading (NLO) simulation in future analyses should improve both of these points.
Furthermore, recent theoretical advances~\cite{topnnlo} allow the calculation of these observables at next-to-next-to-leading order (NNLO) in perturbation theory, which could be directly compared to particle-level measurements and used in a top-quark mass extraction.

\section{Measurements using theoretically calculable observables}\label{sec:theoclean}
All analyses presented so far use Monte Carlo simulation to quantify the dependence of their observables on the top-quark mass, as do the more precise standard measurements.
A different, and theoretically more appealing approach is to use observables whose mass dependence can be calculated beyond leading order in perturbative expansion.
This allows the extraction from data of a mass parameter that is precisely defined in QCD theory.

\subsection{Inclusive production cross section}\label{subsec:crossec}
\begin{figure}
\hfill
\begin{minipage}{0.57\linewidth}
\centerline{\includegraphics[width=1.0\linewidth]{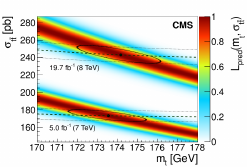}}
\end{minipage}
\hfill
\begin{minipage}{0.42\linewidth}
\centerline{\includegraphics[width=1.0\linewidth]{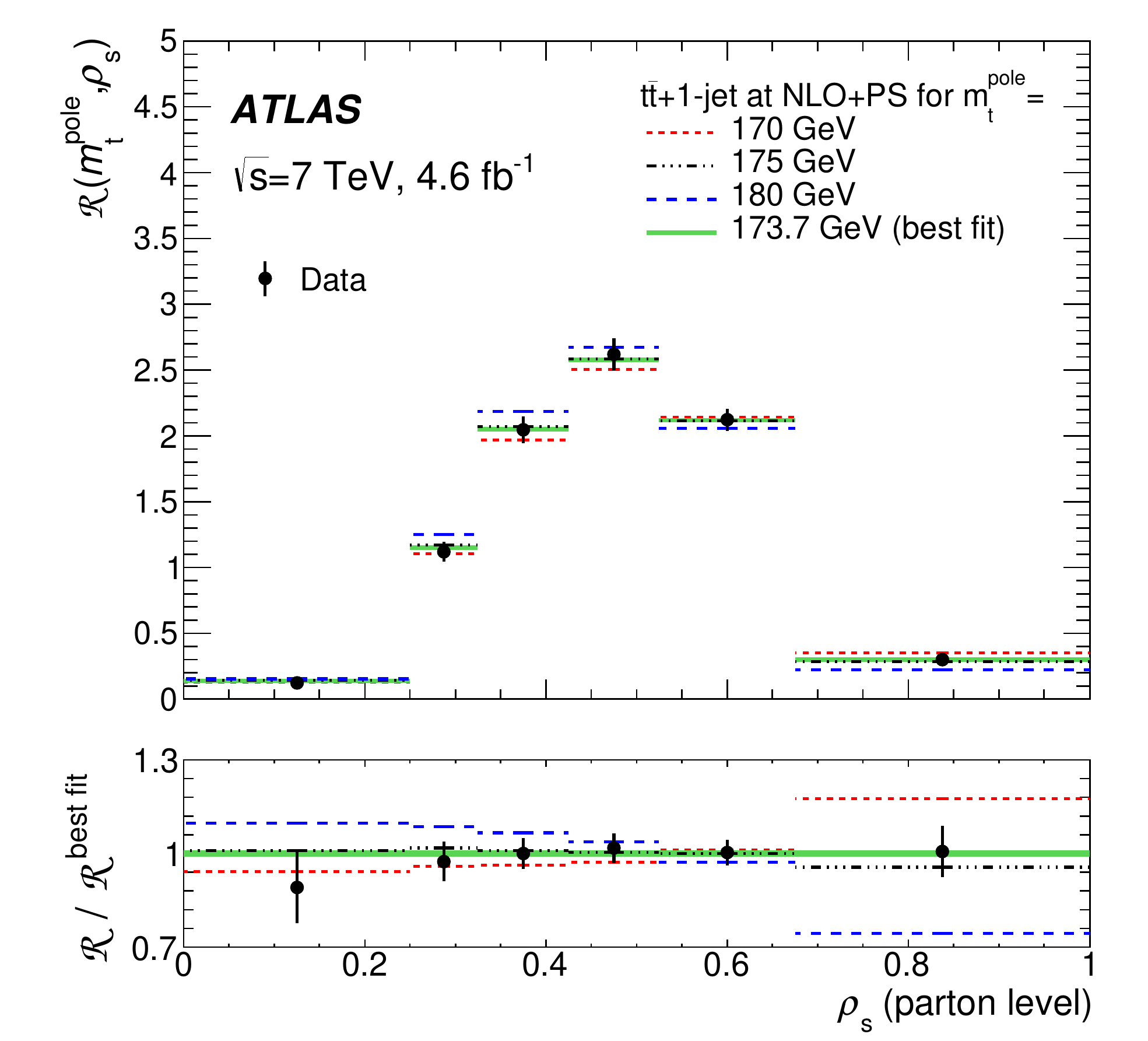}}
\end{minipage}
\caption[]{
  Left: Predicted dependence of the inclusive \ttbar\ production cross section on the top quark pole mass at NNLO, and measured cross section, for 7 and 8~TeV center of mass energy~\cite{cmspole}.
  Right: Observed $\rho_S$ distribution in \ttbar+1 jet events, compared to shapes at several top-quark mass values predicted at NLO~\cite{atlasttj}.
}
\label{fig2}
\end{figure}

The cross section to inclusively produce \ttbar\ pairs can be calculated at NNLO in QCD as a function of the top quark pole mass.
Comparing the prediction at a given center-of-mass energy of 7 or 8~TeV with the corresponding cross section measurements, and taking into account a slight dependence of the detector acceptance on the assumed top-quark mass, yields a direct measurement of the top quark pole mass.
Both ATLAS and CMS have performed such analyses, recently reaching a precision of below 2~GeV for the first time~\cite{cmspole,atlaspole}.
Figure~\ref{fig2}, left, shows the measured production cross section as a function of the assumed top-quark mass, compared with the mass-dependent prediction at NNLO, for the two center-of-mass energies of run I of the LHC.\@

\subsection{\ttbar+1 jet invariant mass}\label{subsec:ttj}
Beyond inclusive cross sections, several kinematic shapes could possibly be exploited to extract a well-defined top-quark mass.
One of the first to be proposed by theorists~\cite{fusterttj} and recently experimentally tested by both the ATLAS and CMS experiments, concerns the primary QCD radiation of the \ttbar\ system~\cite{atlasttj,cmsttj}.
Selecting \ttbar\ events with an additional hard radiation and forming the invariant mass of the \ttbar\ system and the hardest additional jet, an observable ($\rho_S = m_0/m(\ttbar,j)$) can be constructed whose distribution is calculable at NLO and depends on the top-quark mass.
Figure~\ref{fig2}, right, shows the observed distribution of $\rho_S$ unfolded to particle level, compared to predictions at different top-quark masses, calculated at NLO using POWHEG.\@
Unlike the inclusive cross section measurements, these analyses are not limited by beam-related uncertainties and have the potential to surpass the former in precision.
Currently, they are limited by uncertainties stemming from scale variations in the MC used for the unfolding of detector effects and for the calculation of the NLO shape prediction.

\section{Measurements in alternative topologies}\label{sec:topo}
Finally, both CMS and ATLAS perform analysis in event topologies where top quarks are predominantly produced singly in electroweak-mediated interactions~\cite{atlassingle,cmssingle}.
With the current event samples and systematic precision, the potential differences in sensitivity to the hard scattering and the modeling of underlying events and color reconnection flows are so far not observed and the measurements are limited by jet-energy calibration uncertainties.
Nevertheless, the analyses can provide a statistically independent determination of the top-quark mass and have an impact in future combinations.

\section{Conclusion}\label{sec:conclusion}
\begin{figure}
\centering
\includegraphics[width=0.52\linewidth]{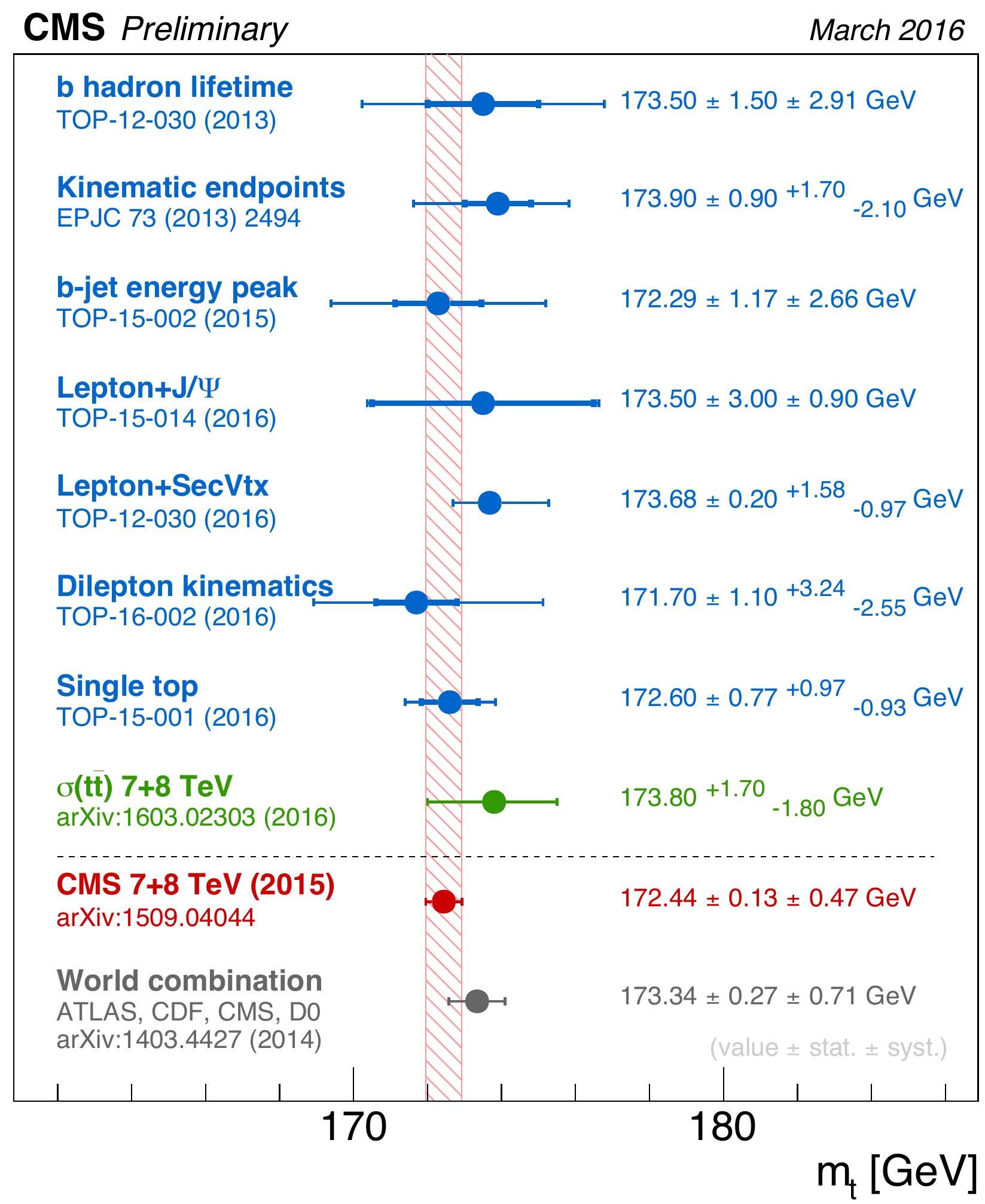}
\caption[]{
	Recent alternative top-quark mass measurements from the CMS collaboration, many of which were presented for the first time at this conference, compared to the combination of standard measurements from CMS and to the 2014 world average~\cite{cmssummaryfigs}.
}
\label{fig3}
\end{figure}

A rich spectrum of alternative approaches to top-quark mass measurements---a summary of recent CMS results is shown Fig.~\ref{fig3}---is being carried out by the ATLAS and CMS collaborations, while standard measurements have reached an unprecedented precision of below 500~MeV.
Alternative mass measurements can on the one hand provide insights in the modeling of top quark events and b-quark hadronization, and on the other hand produce important cross checks by using different mass definitions.
With the expected increased size of datasets during run II of the LHC, many of these will further gain importance and might prove to be crucial in improving the overall precision on the top-quark mass.

\end{document}